\documentclass[preprint,aps,floats,amssymb,showpacs]{revtex4}
\usepackage{epsfig}
\newcommand{\be}{\begin{equation}}
\newcommand{\ee}{\end{equation}}
\newcommand{\bea}{\begin{eqnarray}}
\newcommand{\eea}{\end{eqnarray}}
\newcommand{\bml}{\begin{mathletters}}
\newcommand{\eml}{\end{mathletters}}

\begin{document}

\preprint{IUB-TH-044}

%\twocolumn[\hsize\textwidth\columnwidth\hsize\csname @twocolumnfalse\endcsname

%%%%%%%%%%%%%%%%%%%%%%%%%%%%%%%%%%%%%%%%%%%%%%%%%%%%%%%%%%%%%%%%%%%%%%%%%%

%\wideabs{                       % Uncomment this line for two-column output

\title{Solitons on nanotubes and fullerenes as solutions of a modified
non-linear Schr\"odinger equation   }
\renewcommand{\thefootnote}{\fnsymbol{footnote}}
\author{Yves Brihaye\footnote{Yves.Brihaye@umh.ac.be}}
\affiliation{Facult\'e des Sciences, Universit\'e de Mons-Hainaut,
7000 Mons, Belgium}
\author{Betti Hartmann\footnote{b.hartmann@iu-bremen.de}}
\affiliation{School of Engineering and Sciences, International University
Bremen (IUB), 28725 Bremen, Germany}

\date{\today}
\setlength{\footnotesep}{0.5\footnotesep}

%%%%%%%%%%%%%%%%%%%%%%%%%%%%%%%%%%%%%%%%%%%%%%%%%%%%%%%%%%%%%%%%%%%%%%%%%%
\begin{abstract}
Fullerenes and nanotubes consist of a large number of carbon atoms
sitting on the sites of a regular lattice. For pratical reasons it is often
useful to approximate the equations on this lattice in terms of
the continuous equation. At the moment, the best candidate for such an
equation is the modified non-linear Schr\"odinger equation.
In this paper, we study the modified non-linear Schr\"odinger equation, which
arises as continuous equation in a system describing an excitation
on a hexagonal lattice. This latter system can e.g.
describe electron-phonon interaction on fullerene related structures
such as the buckminster fullerene and nanotubes. Solutions of this modified
non-linear Schr\"odinger equation, which have solitonic character,
can be constructed using an Ansatz for the wave function, which has time
and azimuthal angle dependence introduced previously in the study of spinning
boson stars and Q-balls. We study these solutions on a sphere, a disc and on a cylinder. 

Our construction suggests that non-spinning as well as spinning solutions, which have
a wave function with an arbitrary number of nodes, exist.
We find that this property is closely related to the series of well-known
mathematical functions, namely the Legendre functions when studying the
equation on a sphere, the Bessel functions when studying
the equation on a disc and the trigonometric functions, respectively,
when studying the equation on a cylinder.
\end{abstract}

\maketitle 
\renewcommand{\thefootnote}{\arabic{footnote}}
\section{Introduction}
Fullerenes and fullerene related structures have gained a lot of interest
in the past 20 years. Fullerenes were discovered for the first time in
1985 \cite{cks} and  are carbon-cage molecules with a large number $n$ of
carbon (C) atoms bonded in a nearly spherically symmetric configuration.
The configuration is such that three of the valence electrons of the carbon atom
form the bounds with the neighbouring carbon atoms. This means, that one ``free''
valence electron can ``hop'' along the the $n$ positions in the fullerene.

The $C_{60}$ fullerene, the so-called ``buckminster'' fullerene, has 60 carbon
atoms arranged on a lattice with 20 hexagons and 12 pentagons. It has a diameter of
 $d_{60}\approx 7$ \AA. In general, fullerenes with $n$ carbon atoms
consists of lattices with 12 pentagons and  $(\frac{n}{2}-10)$ hexagons, where $n\geq 24$ has to be even.

Fullerides are alkali-doped fullerenes, i.e. fullerenes on which alkali metal atoms
such as rubidium (Rb), potassium (K), caesium (Cs) or sodium (Na) are put.
These alkali metals donate one electron each. It was observed that such ``doping''
leads to a metallic or even superconducting behaviour \cite{haddon}. The interesting
thing about this is that the transition temperature $T_c$ of the fullerides
is quite high for superconductors, e.g. 
$T_c=33$ K for a RbCs$_{2}$C$_{60}$ \cite{tanigaki}.
In an attempt to explain this high transition temperature, it was found that
phonon-electron interactions in the fullerides can explain this phenomenon 
quite well \cite{varma}.

Nanotubes are fullerene related structures, which were discovered for the first time
in 1991 \cite{iijima}. These molecules have cylindrical shape and
the carbon atoms are arranged on a hexagonal grid. The ends
of the nanotubes are typically closed by caps of pentagonal rings. While the
tubes discovered in 1991 were multi-layered, single-walled tubes were synthesised
in 1993 \cite{iijima2}.
The tubes are distinguished 
by their chirality, diameter and lengths. They have average diameters of 1.2 to 1.4 nm
and can be up to 2 mm long. Concerning their chirality, the tubes are typically
put into three different classes: armchair, zigzag or chiral tubes \cite{nanotubes}.
Depending on their structure, the tubes have different mechanical, thermal, optical
and electrical properties. The chirality determines e.g. whether  the tubes
behave metallic or semiconducting.
These facts also lead to the conclusion that a distortion of the hexagonal
lattice would influence the energy-band gap. A distortion can be achieved
in two ways: either by external, mechanical forces such as bending, stretching or twisting
or by an interaction of the lattice with an internal excitation. Trying to
explain the dispersion-free energy transport in biopolymers, 
Davydov realised \cite{davydov} that the interaction of an internal excitation with the
distorted lattice (whose distortion was initially caused by the internal
excitation) leads to a localised state, a soliton \cite{scott}.
For the case of biopolymers, the internal excitation is an amide I vibration.

Motivated by the research on fullerene related structures, 
the interaction of a 2-dimensional hexagonal lattice with an excitation caused
by an ``excess electron'' was studied recently \cite{hz}.
The Hamiltonian is of Fr\"ohlich-type and describes ``electron-phonon'' interactions.
It was found that the existence of soliton-like structures depends crucially on the
value of the ``electron-phonon'' coupling.

Interestingly, in the continuum limit, this latter system of equations reduces to
a modified non-linear Schr\"odinger equation, where the extra term
appearing due to the discreteness of the lattice and the
``electron-phonon'' interaction, leads to the stabilisation of the
soliton. The appearance of this type of equation has been observed  also
for a similar system on a quadratic lattice \cite{bepz,bpz}. Consequently, the
modified non-linear Schr\"odinger equation was studied on a 2-dimensional
plane in \cite{bhz}  and on a sphere in \cite{bh} using an Ansatz previously introduced
in the study of boson stars \cite{bs} and Q-balls \cite{volkov}. Non-spinning as well
as spinning solutions with and without nodes were constructed. 

In this paper, we study the modified non-linear Schr\"odinger equation arising as
continuum limit of the equations describing the interaction of an electron-like excitation
with a hexagonal lattice. In Section II, we review some
generalities. Using the Ansatz introduced in \cite{bs,volkov}, we 
construct solutions for three different cases, which we
describe in Sections III-V. In Section III, we discuss the results
for the solutions on the sphere obtained in \cite{bh} from another point of view, namely
we give the spectrum of the equation. In Section IV, we discuss the solutions
on a 2-dimensional disc with fixed radius and demonstrate how the solutions
obtained on the plane \cite{bhz} can be understood in this context.
In Section V, we discuss the equation on a cylinder with fixed radius and
fixed length.
We give our conclusions in Section VI.

%%%%%%%%%%%%%%%%%%%%%%%%%%%%%%%%%%%%%%%%%%%%%%%%%%%%%%%%%%%%%%%%%%%%%%
\section{The Modified Non-linear Schr\"odinger equation: Generalities}
%%%%%%%%%%%%%%%%%%%%%%%%%%%%%%%%%%%%%%%%%%%%%%%%%%%%%%%%%%%%%%%%%%%%%%
The starting point is a coupled systems of discrete equations
describing the interaction of an excitation (e.g. an ``excess electron'')
with a regular lattice in two dimensions. The excitation is given
in terms of a complex valued scalar field $\psi$, while the lattice displacements 
are described by two fields, one standing for the displacement in $x$, 
the other for that
in $y$ direction. These discrete equations were studied for the quadratic
lattice in \cite{bepz,bpz} and for the hexagonal lattice in \cite{hz}.
Remarkably, it was found
for the hexagonal lattice that in the stationary limit it is possible
to replace the full system of equation, in which the excitation field $\psi$
and the displacement fields are coupled, by a modified discrete
non-linear Schr\"odinger equation. This is not possible for the quadratic lattice.

In the continuum limit, both systems of equations reduce to a modified
non-linear Schr\"odinger equation. The new term appearing in this
equation, which results from
both the discreteness of the lattice as well as from the interaction
between the lattice and the 
complex valued scalar field $\psi$ (``lattice-excitation interaction''),
leads to the
stabilisation of the solution.
The difference between the quadratic and
the hexagonal case is just the  coefficient in front of  the non-linear 
term in the equation, which, however is of the same order of magnitude
for the two cases. 

Since we are dealing with fullerene related structures here, we consider the
modified non-linear Schr\"odinger equation, which arises as continuum
limit in the hexagonal case \cite{hz, bhz}.
In dimensionless variables it reads:
\begin{equation}
\label{cnls}
i\frac{\partial \psi}{\partial t}+\Delta \psi +4g\psi\left(\vert\psi\vert^2+
\frac{1}{8}\Delta \vert\psi\vert^2\right)=0 \ .
\end{equation}
 $g$ is the coupling constant of
the system
and determines the ``strength'' of the non-linear character of the equation.
Especially, as can be seen from the derivation of the modified non-linear
Schr\"odinger equation from the discrete equations \cite{hz}, the mass
of the carbon atom is encoded into $g$.

%%%%%%%%%%%%%%%%%%%%%%%%%%%%%%%%%%%%%%%%%%%%%%%%%%%%%%%%%
\section{Equation on the sphere}%%%%%%%%%%%%%%%%%%%%%%
%%%%%%%%%%%%%%%%%%%%%%%%%%%%%%%%%%%%%%%%%%%%%%%%%%%%%%%%%
In this section, we will discuss the equation (\ref{cnls}) above
on a 2-dimensional sphere of radius $R_s$ aiming to describe ``electronic''
excitations on fullerenes and fullerides.

The Laplacian operator in this case reads:
\begin{equation}
\Delta_{sphere}=\frac{1}{R_s^2}\left(\frac{\partial^2}{\partial\theta^2}+
\cot\theta\frac{\partial}{\partial\theta}+
\frac{1}{\sin^2\theta}\frac{\partial^2}{\partial\varphi^2}\right)  \ ,
\end{equation}
where $\theta$ $\epsilon$ $[0:\pi]$
and $\varphi$ $\epsilon$ $[0:2\pi]$. Note that 
$\psi=\psi(t,\theta,\varphi)$ is a complex valued scalar field.

Solutions of (\ref{cnls}) considered on the sphere with radius $R_s$ 
can be characterised by their norm:
\begin{equation}
\eta^2=\int\limits_0^{2\pi}\int\limits_0^{\pi} |\psi|^2 \sin\theta d\theta d\varphi
 \end{equation}
 as well as by their energy:
 \begin{equation}
 E=\int\limits_0^{2\pi}\int\limits_0^{\pi} \left( |\vec{\nabla}\psi|^2-2g |\psi|^4 + 
\frac{1}{4}g(\vec\nabla |\psi|^2)^2\right) \sin\theta d\theta d\varphi \ .
 \end{equation}
We will construct normalised solutions, i.e. $\eta^2=1$. Note that
in \cite{bh} a different strategy was employed. By 
rescaling the wave-function  $\psi\rightarrow\psi/\sqrt{g}$, the coupling
constant $g$ can be eliminated from the equation. (Of course,
the normalisation of the wave-function is lost in this approach, but 
since $\eta$ was plotted, the corresponding value of $g$ could be easily
determined.)
In this paper, we put the emphasis on the evolution of the spectrum
(corresponding to normalised eigenfunctions) as a function of the
coupling constant $g$. Note that $g$ determines the ``strength''
of the non-linear part of the equation and that for $g=0$, the equation
becomes the ordinary Schr\"odinger equation on the sphere, which, of course,
is linear.

%%%%%%%%%%%%%%%%%%%%%%%%%%%%%%%%%%%%%%%%
\subsection{Discussion of the solutions}
%%%%%%%%%%%%%%%%%%%%%%%%%%%%%%%%%%%%%%%%%
To construct explicit solutions, we employ the following axially symmetric
Ansatz \cite{bs,volkov}:
\begin{equation}
\label{ansatz}
\psi(t,\theta,\varphi)=e^{-i\omega t+i m \varphi } \Phi(\theta) \ .
\end{equation}
The solutions with $m=0$ are non-spinning solutions, while the spinning
solutions will have $m\neq 0$.

Inserting this Ansatz into (\ref{cnls}), we find the following equation \cite{bh}:
\begin{equation}
\label{phi}
\Phi''+\cot \theta \Phi' +
\Phi'^2\frac{g\Phi}{1+g\Phi^2}+\frac{4g\Phi^3+\omega\Phi}{1+g\Phi^2}R_s^2 
- \frac{m^2}{1+g\Phi^2} \frac{1}{\sin^2 \theta } \Phi=0  \ ,
\end{equation}
where the prime denotes the derivative with respect to $\theta$.

This is a priori a non-linear equation with external parameters $g$ and $\omega$.
Here, we will consider it from a slightly different point of view.
Let us rewrite (\ref{phi}) as follows:
\begin{equation}
\label{nlsemod}
-\left(\Phi''+\cot \theta \Phi' +\Phi'^2\frac{g\Phi}{1+g\Phi^2}+
\frac{g\Phi^3 R_s^2 (4-\omega)}{1+g\Phi^2} - \frac{m^2}{1+g\Phi^2} \frac{1}
{\sin^2 \theta } \Phi\right)= \Omega \Phi
\end{equation}
where
\begin{equation}
\Omega\equiv\omega R_s^2 
\end{equation}
is the effective spectral parameter.
The equation then appears as an eigenvalue Sch\"odinger equation
supplemented by a non-linear term coupled with the parameter $g$.
The solutions in the linear case, i.e. for $g=0$ are well-known
and it is very likely that - by continuity- their deformation due to the
non-linear term can be constructed at least for small values of $g$.
However, this does not exclude the occurence of supplementary
branches of solutions, which may appear typically due to the
non-linear structure of the equation.

In order to construct regular solutions, (\ref{nlsemod}) has to be solved
subject to the following boundary conditions:
\begin{equation}
\label{bc1}
\Phi'|_{\theta=0}=\Phi'|_{\theta=\pi}=0 \ \ \ {\rm for} \ \ m=0
\end{equation}
and
\begin{equation}
\label{bc2}
\Phi(\theta=0)=\Phi(\theta=\pi)=0 \ \ \ {\rm for} \ \ m\neq 0  \ .
\end{equation}

To solve the equation (\ref{nlsemod}) numerically subject to the boundary
conditions (\ref{bc1}), respectively (\ref{bc2}) for generic
values of $g$ and $R_s$, we use a numerical routine based 
on the Newton-Raphson method \cite{colsys}. Our strategy
is to use $\Phi(\theta=0)$, respectively $\Phi'|_{\theta=0}$ 
as a ``shooting'' parameter for the case $m=0$ and $m\neq 0$.

%%%%%%%%%%%%%%%%%%%%%%%%%%%%%%%%%%%%%%%%%%
\subsubsection{The case $g=0$}
%%%%%%%%%%%%%%%%%%%%%%%%%%%%%%%%%%%%%%%%%%
In the case $g=0$, (\ref{nlsemod}) reduces to the well-known 
Legendre equation
\begin{equation}
\Phi'' + \cot\theta \Phi' - m^2 \frac{1}{ \sin^2\theta} \Phi + \Omega \Phi=0 \ .
\end{equation} 
The solutions of this equation are the associated Legendre functions: 
\begin{equation}
\Phi(\theta)=P_l^m (\cos\theta) \ \ \ {\rm with} \ \ \Omega = l(l+1) \ ,
\end{equation}
where
\begin{equation}
l \in \mathbb{N} \ \ , \ \  m=-l, -l+1, ..., l-1, l  \ . 
\end{equation}
The spatial part of (\ref{ansatz}) are then the 
spherical harmonics $Y^l_m(\theta,\varphi)=e^{im\varphi}P^l_m(\cos\theta)$. 
Notice, however, that in the equation studied here there is no distinction
between $m$ and $-m$. From now on, we will assume $m$ to be positive,
but it is understood that the solution with corresponding $-m$
can be trivially obtained. 

The spherical harmonics have a well-defined parity, i.e.
they are either symmetric or anti-symmetric under the reflection
$\Phi(\theta) \rightarrow \Phi(\pi-\theta)$. In addition, the number
of zeros in the interval $]0:\pi[$ is equal to $ l-|m|$.

These results are, of course, very well- known, but we mention them for
completeness and because the spectrum of the
full equation will approach the spectrum of the Legendre equation
in the $g\rightarrow 0$ limit.

%%%%%%%%%%%%%%%%%%%%%%%%%%%%%%%%%%%%%%%%%%
\subsubsection{Deformed Legendre functions}
%%%%%%%%%%%%%%%%%%%%%%%%%%%%%%%%%%%%%%%%%%
All the solutions discussed in the previous section are 
continuously deformed when the parameter $g$ is chosen to be larger than zero.
For every available spherical harmonic, we observe that
a branch of solutions in $g$ exists, on which the solutions have the
same symmetry than in the $g=0$ limit. Our numerical results
further indicate that the number of zeros $k$ of the solution is
preserved all along the branch and is equal to $k\equiv l-m$. 

Now, we will discuss in detail the numerical solutions corresponding to
the deformations of the spherical harmonics $Y^0_0$, $Y_0^1$ and
$Y_1^1$.  

\begin{figure}
\centering
\epsfysize=20cm
\mbox{\epsffile{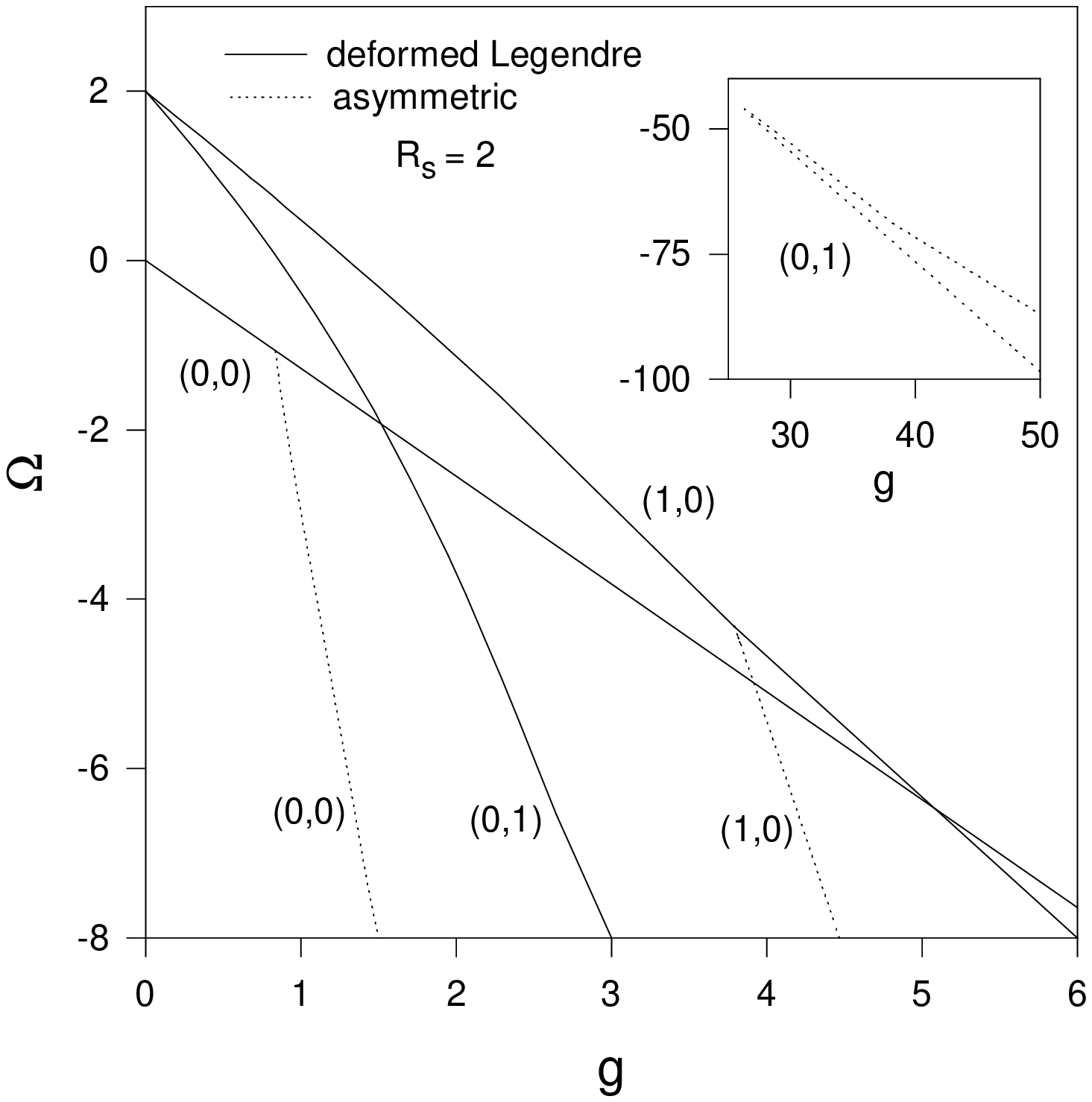}}
\caption{\label{fig1}
The eigenvalue $\Omega=\omega R_s^2$ is shown as a function of $g$
for $(m,l)=(0,0)$, $(0,1)$ and $(1,0)$. We have chosen $R_s=2$.
The solid lines denote the deformed Legendre functions, while the dotted
lines are the asymmetric solutions (see text for more details).}
\end{figure}

\begin{itemize}
\item
The non-spinning ($m=0$) and 
nodeless ($l=0$, $k=0$) solution corresponding to the case $Y^0_0$
is in fact
the constant solution of the equation, namely
\begin{equation}
\Phi=\sqrt{\frac{-\omega}{4g}}   \ .
\end{equation}
The eigenvalue $\Omega$ of the corresponding
normalised solution has a linear dependence of the parameter $g$:
\begin{equation}
\Omega=-\frac{R_s^2 g}{\pi}  \ .
\end{equation} 
This dependence is shown in Fig.\ref{fig1} for $R_s=2$.

\item  The non-spinning ($m=0$), one-node ($l=1$, $k=1$) 
solution is anti-symmetric
under the reflection $\theta\rightarrow \pi-\theta$.
The function $Y_0^1 \propto \cos\theta$ is progressively deformed
by the non-linear term for $g > 0$. For $g\approx 10$ the profile of the
function $\Phi(\theta)$ possesses a plateau $\Phi=0$ surrounding the
region of $\theta=\pi/2$. Away from this plateau, the solutions
resemble two disconnected solitons located at the two poles of the
sphere, i.e. the maxima of $\Phi^2(\theta)$ are situated
around $\theta=\pi$ and $\theta=0$. Stated in other words, the probability
density vanishes in a region surrounding the equator, while it has
local maxima at the two poles. Let us
stress, that the physical parameters remain
monotonic functions of the ``shooting'' parameter $\Phi(\theta=0)$.

The dependence of $\Omega$ on $g$ is shown in Fig.\ref{fig1} for $R_s=2$.
For $g=0$, we have $\Omega=2$, as expected. It is worth noticing
that the two eigenvalues corresponding to the quantum numbers
$m=l=0$ and $m=0$, $l=1$ cross at $g\approx 1.5$. This would
a priori mean, that the ``excited'' solution becomes the ``ground state''
of the equation. However, as we will see in the next section, the appearance
of new asymmetric solutions will turn out to have even lower values
of $\Omega$.

\item The spinning ($m=1$), no-node ($l=1$, $k=0$) solutions
arrive as deformations
of the spherical harmonic $Y_1^1\propto \sin\theta e^{i\varphi}$.
This solution is thus symmetric under the reflection $\theta\rightarrow \pi-\theta$.
In contrast to the non-spinning solutions, we notice here
that when the coupling constant $g$ is of order
$2.5$, the structure of the solutions becomes very sensitive
to the value of the ``shooting'' parameter $\Phi'|_{\theta=0}$.
In other words, there exists a critical value of  $\Phi'|_{\theta=0}\equiv
\Phi'_{(cr)}|_{\theta=0}$
(depending on $R_s$) such that no solutions exist for
$\Phi'|_{\theta=0} >
\Phi'_{(cr)}|_{\theta=0}$. Our numerical results, however,
suggest that solutions of arbitrary large values of $g$
can be constructed when considering the limit $\Phi'|_{\theta=0} \rightarrow
\Phi'_{(cr)}|_{\theta=0}$. 
The corresponding eigenvalue of $\Omega$ 
decreases monotonically with $g$ from $\Omega=2$ (for the case $g=0$)
as can be seen in Fig.\ref{fig1} for the case $R_s=2$. Of course, this curve
also represents the solution with  $m=-1$, so that the eigenvalue $\Omega$
is three-fold degenerated in the limit $g=0$.  

Finally, let us state that for these solutions, 
the function $\Phi$ has a maximum at the equator and vanishes at the poles. 
This is an interesting result with view to \cite{harigaya}.
In this latter paper it was found that if a C$_{60}$ or a C$_{70}$ molecule
is doped with one or two excess electrons, the additional
charges accumulate nearly along an equatorial line of the molecule.

\end{itemize}
 
Of course, deformations of the Legendre functions for higher $m$ and
$l$ can be constructed in a systematic way.

%%%%%%%%%%%%%%%%%%%%%%%%%%%%%%%%%%%%%%%%%%
\subsubsection{Asymmetric solutions}
%%%%%%%%%%%%%%%%%%%%%%%%%%%%%%%%%%%%%%%%%%
All solutions of the non-linear equation (\ref{nlsemod})
discussed previously appear as continuous deformations
of the solutions available in the linear limit $g=0$.
However, in this section, we will exhibit
new branches of solutions, which are fundamentally non-linear
phenomena and as these have no linear counterparts.
Numerically, we find that these new branches of solutions
exist for sufficiently high values of the
coupling constant $g$ parametrising the non-linear effect and
that some of these branches bifurcate into the branches of the
deformed Legendre functions at critical values of $g=g^{(m,l)}_{cr}(R_s)$, 
which depend on $l$, $m$ 
(and thus on the number of nodes $k=l-m$) as well
as on the radius $R_s$.

These solutions are charaterised by the fact that
they are asymmetric with respect to the reflection 
$\theta\rightarrow \pi-\theta$.

%%(ii) For the values of $g$ that these solutions exist, they have
%%lower eigenvalue $\Omega$ than the corresponding
%%deformed Legendre function with the same values of $m$, $l$ and $g$.

\begin{figure}
\centering
\epsfysize=20cm
\mbox{\epsffile{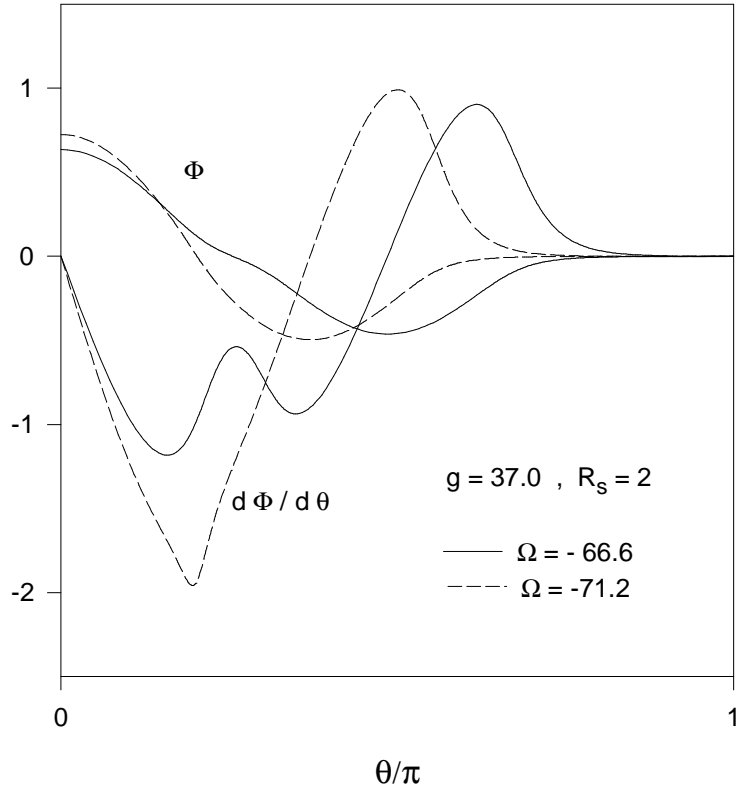}}
\caption{\label{fig2}
The profiles of the function $\Phi(\theta)$ and the derivative $\frac{d\Phi}{d\theta}$
are shown for $(m,l)=(0,1)$ (non-spinning, nodeless solutions),
$g=37$, sphere-radius $R_s=2$ and two different values of the spectral parameter $\Omega=\omega R_s$.}
\end{figure}

\begin{itemize}
\item Asymmetric, non-spinning ($m=0$) and nodeless ($l=0$, $k=0$) solutions exist
for $g > g^{(0,0)}_{cr}(R_s=2)\approx 0.838$.
 When the parameter $g$ is chosen larger
than this critical value, our numerical results indicate
that this solution has a maximum at $\theta=0$ and a minimum at
$\theta=\pi$ and decreases monotonically between these two extrema, which
are strictly positive.
Thus, the solution is neither symmetric nor anti-symmetric under the
reflection $\theta\rightarrow \pi-\theta$. Instead, this reflection
applied to the solution leads to an equivalent solution
with a maximum (respectively minimum) at $\theta=\pi$ ($\theta=0$).
The solutions thus represent a physical state with probability density
concentrated in one of the hemispheres. This is a new feature that
cannot arise in the deformation of the Legendre functions.

In the limit $g\rightarrow g^{(0,0)}_{cr}$ the values of the maximum
and the minimum approach each other and the solution becomes constant.
The branch of these asymmetric solutions bifurcates
into the branch of constant solutions (deformed $Y_0^0$) described
in the previous section. The corresponding value of $\omega$ of these asymmetric
solutions is lower than the one of the symmetric branch.
This is shown
in Fig.\ref{fig1} (dotted line) for $R_s=2$.

For generic values of $R_s$, the bifurcation of the asymmetric
branch occurs at
\begin{equation}
g^{(0,0)}_{cr}(R_s)=\frac{\pi}{R_s^2-\frac{1}{4}}  \ \ ,
\end{equation}
with
\begin{equation}
\Omega_{cr}^{(0,0)}=-\frac{R_s^2}{R_s^2-1/4}  \ .
\end{equation}

\item Asymmetric, non-spinning ($m=0$) and one-node ($l=1$, $k=1$)
solutions exist for $g > g^{(0,1)}_{cr}(R_s=2)\approx 25$. In this region
of parameter values, we find that in fact two branches of solutions exist
(see Fig.\ref{fig1}). For a fixed value of $g$, the solution with the lower
value of $\Omega$ has a bigger value of $\Phi(0)$ and its minimum
lies at smaller values of $\theta$. This is demonstrated for $g=37$, $R_s=2$
in Fig.\ref{fig2}.
The values of  $\omega$ corresponding to these asymmetric solutions
are higher than the values of the corresponding 
(i.e. with for the same values of $g$) antisymmetric solution on the branch
of deformed Legendre functions $Y_0^1$. In contrast to the other described cases,
we have not succeeded to construct another (a third)
branch which would eventually bifurcate from the
branch of deformed Legendre functions.

%the numerical analysis turns out to be involved
%and it was difficult to produce the full asymmetric branch
%as well as to determine the critical value of $g$ to a better
%accurancy. The reason for this is that both the anti-symmetric (deformed
%Legendre) and the asymmetric solutions have profiles which are
%nearly equal on a big part of the interval of $\theta$
%(typically $\theta\in [0:0.8]$).
%Only for $\theta > 0.8$ do these solutions deviate significantly.
%We have not attempted to perform a more detailed analysis of the
%branch of asymmetric solutions.

\item  Asymmetric, spinning ($m=1$) and nodeless ($l=1$, $k=0$) solutions
exist for $g > g^{(1,1)}_{cr}(R_s=2)\approx 3.79$. Solving the
equations for values of $g$ larger than this critical value,
we were able to construct a branch of asymmetric solutions
(and the corresponding mirror symmetric one). The maximum of these
solutions is located  between $\theta=0$ and $\theta=\pi/2$ such that
the probability density is located in one of the hemispheres.
As before, the frequency of the asymmetric solution is lower
than the frequency of the corresponding (i.e. same $g$) solution
on the deformed $Y_1^1$ branch.
\end{itemize}

The construction of asymmetric branches with higher values
of the spin $m$ and/or higher node numbers $k$ becomes very tricky.
This is mainly due to the occurence of many solutions existing
for the same values of the parameters and also due to the fact that the profiles
of the
symmetric and asymmetric (or anti-symmetric and asymmetric) solutions
can be very close to each other on a large part
of the $[0: \pi]$ interval. The construction of asymmetric branches
can thus not be achieved systematically.

%%%%%%%%%%%%%%%%%%%%%%%%%%%%%%%%%%%%%%%%%%%%%%%%%%%%%
\section{Equation on the disc}
%%%%%%%%%%%%%%%%%%%%%%%%%%%%%%%%%%%%%%%%%%%%%%%%%%%%%
In \cite{bhz} the equation (\ref{cnls}) was discussed on a two-dimensional
plane and non-spinning as well as spinning solutions with nodes were
constructed. Here, we will treat this problem differently. With view to
the discussion of the equation on a sphere, we will discuss here
the equation on a two-dimensional disc with radius $R_d$ parametrized by the
coordinates $\rho$ and $\varphi$. The case studied in \cite{bhz}
then arises as limit $R_d\rightarrow \infty$. The Laplacian
is given by:
\begin{equation}
\Delta_{disc}=\frac{\partial^2}{\partial \rho^2}+\frac{1}{\rho}\frac{\partial}{\partial \rho} + \frac{1}{\rho^2}\frac{\partial^2}{\partial \varphi^2} 
\end{equation}
with $\varphi\in [0:2\pi]$. The norm reads:
\begin{equation}
\eta^2=\int\limits_0^{2\pi} 
\int\limits_0^{R_d} |\psi|^2 \rho d\rho d\varphi \ 
\end{equation}
and the energy is given by:
 \begin{equation}
 E=\int\limits_0^{2\pi}\int\limits_0^{R_d} \left( |\vec{\nabla}\psi|^2-2g |\psi|^4 + 
\frac{1}{4}g(\vec\nabla |\psi|^2)^2\right) \rho d\rho d\varphi \ .
 \end{equation}

%%%%%%%%%%%%%%%%%%%%%%%%%%%%%%%%%%%%%%%%%
\subsection{Discussion of the solutions}
%%%%%%%%%%%%%%%%%%%%%%%%%%%%%%%%%%%%%%%%%
As for the case on the sphere, we adopt the following Ansatz for the complex valued
function $\psi(t,\rho,\varphi)$:
\begin{equation}
\psi(t,\rho,\varphi)=e^{-i\omega t+im\varphi} \phi(\rho) \ .
\end{equation}
Inserting this Ansatz gives the following equation:
\begin{equation}
\phi''+\frac{\phi'}{\rho} + \frac{g \phi \phi'^2}{1+g\phi^2} + \frac{
4g \phi^3+\omega \phi}{1+g\phi^2} - \frac{m^2 \phi}{\rho^2 (1+g\phi^2)}=0 \ ,
\end{equation}
where the prime now denotes the derivative with respect to $\rho$.

We can rewrite, analogue to before, the equation in the following way:
\begin{equation}
\label{eqdisc}
-\left(\phi''+\frac{\phi'}{\rho} + \frac{g \phi \phi'^2}{1+g\phi^2} 
+ \frac{
g \phi^3(4-\omega)}{1+g\phi^2} - \frac{m^2 \phi}{\rho^2 (1+g\phi^2)}\right)=
\omega\phi  \ .
\end{equation}
This equation looks like the one studied in \cite{bhz}, however, we impose
different boundary conditions here:
\begin{equation}
\label{bc3}
\partial_{\rho} \phi|_{\rho=0}=0 \ \ , \ \ \phi(R_d)=0 \ \ \ {\rm for} \ \ m=0
\end{equation}
and
\begin{equation}
\label{bc4}
\phi(\rho=0)=0 \ \ , \ \ \phi(R_d)=0 \ \ \ {\rm for} \ \ m\neq 0  \ .
\end{equation}
Again, we used the numerical routine described in \cite{colsys} to solve
the equation (\ref{eqdisc}) subject to the boundary conditions 
(\ref{bc3}), respectively (\ref{bc4}).

%%%%%%%%%%%%%%%%%%%%%%%%%%%%%%%%%%%%%%%%%%%%%%%%%%%%%%%%%%%%%%%%%%%%%%%%%
\subsubsection{The $g=0$ limit}
%%%%%%%%%%%%%%%%%%%%%%%%%%%%%%%%%%%%%%%%%%%%%%%%%%%%%%%%%%%%%%%%%%%%%%%%%
In the limit $g=0$ (\ref{eqdisc}) reduces to the linear 
Schr\"odinger equation
in polar coordinates 
\begin{equation}
\label{eqdiscnew}
\phi''+\frac{\phi'}{\rho} - \frac{m^2 \phi}{\rho^2}+
\omega\phi  =0 \ .
\end{equation}
Note that for static solutions, i.e. $\omega=0$, the equation reduces to the
Euler equation with solutions of the form $\psi(t,\rho,\varphi)
\equiv \psi(\rho)\propto
e^{im\varphi} \rho^{ m}$.
For $\omega\neq 0$, the equation becomes the Bessel equation if we define
$x\equiv \sqrt{\vert \omega \vert}\rho$ and $F(x)\equiv \phi(x)$:
\begin{equation}
\frac{d^2 F}{dx^2}+\frac{1}{x}\frac{d F}{dx} 
+ \left(1-\frac{m^2}{x^2}\right) F=0
\end{equation}
The solutions of this equation, which are regular at the origin are of course
the well-known Bessel functions $J_m(x)$ for $\omega > 0$ and the modified
Bessel functions $I_m(x)$ for $\omega < 0$. Remembering that
the Bessel functions  $J_m(x)$ are oscillating and admit an infinite
number of nodes on the positive real axis, it is easy to construct
solutions of (\ref{eqdiscnew}) with $k$ nodes. This is possible assuming that
\begin{equation}
\phi(\rho) = J_m(\sqrt{\omega}\rho)    \ \ {\rm with} \ \     \phi(R_d) = 0
\end{equation}
and choosing the value of $\omega$ in such the way that the 
value $\rho = R_d$ coincides with the $k$-th zero 
(we call it  $x = x_{m,k}$ ) of the Bessel
function $J_m$, i.e.  $\sqrt{\omega} = x_{m,k}/R_d$.

This indicates that in the limit $g=0$ solutions with angular momentum
$m$ and with $k$ nodes exist and the spectrum of the equation is positive.
The corresponding functions  are convex.

%%%\begin{equation}
%%%\label{besselj0}
%%%\phi(r>>1) = J_0(x)
%%%\sim \frac{cos(x-\pi/4)}{\sqrt x} \ \ , \ \ {\rm if} \ \omega < 0
%%%\end{equation}
%%%\begin{equation}
%%%\label{besselk0}
%%%  \phi(r>>1) = K_0(x)
%%% \sim \frac{exp(-x)}{\sqrt{x}} \ \ , \ \ {\rm if} \ \omega > 0
%%% \end{equation}

%%The function $\phi(r)$ therefore oscillates around $\phi = 0$
%%for $\omega < 0$ and decays exponentially  for $\omega > 0$.

%%%%%%%%%%%%%%%%%%%%%%%%%%%%%%%%%%%%%%%%%%%%%%%%%%%%%%%%%%%%%%%%%%%%%%%%%
\subsubsection{The $g\neq 0$ case}
\begin{figure}
\centering
\epsfysize=20cm
\mbox{\epsffile{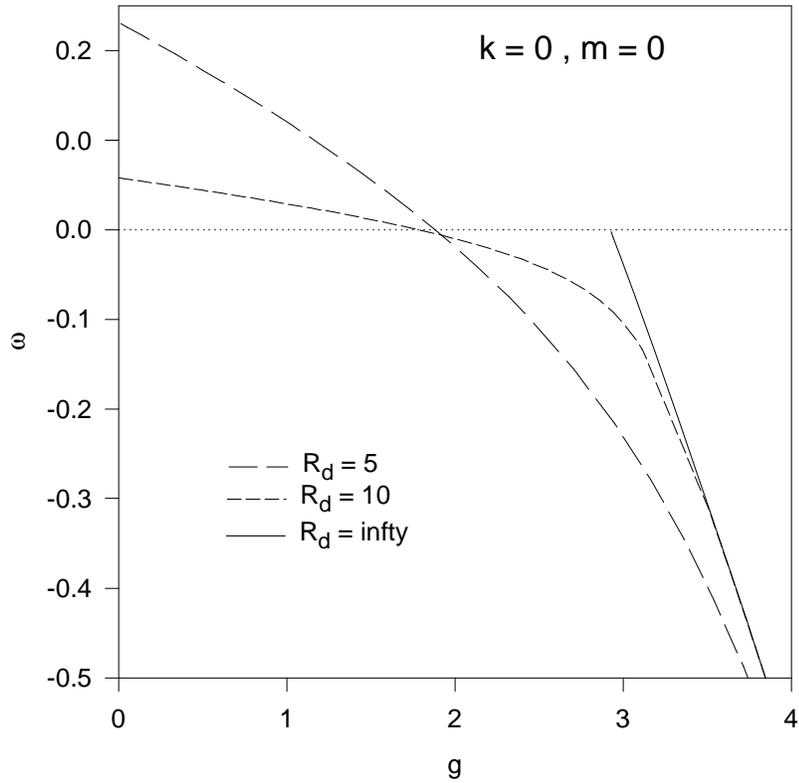}}
\caption{\label{fig3}
The eigenvalue $\omega$ is shown as a function of $g$
for the non-spinning ($m=0$), nodeless ($k=0$) solutions on the disc. 
We have chosen $R_d=5$, $10$ and $\infty$. Note that $R_d=\infty$ corresponds
to the case of the plane.}
\end{figure}

%%%%%%%%%%%%%%%%%%%%%%%%%%%%%%%%%%%%%%%%%%%%%%%%%%%%%%%%%%%%%%%%%%%%%%%%%
If the coupling constant $g$ is positive but small, we expect that
the pattern of solutions will be similar to the one
available in the limit $g=0$ discussed above. Our numerical results indeed
confirm this expectation. 
For small values of $g$, the solutions are still represented by convex functions
and the values of $\omega$ are positive and decrease slowly with $g$
as shown in Fig. \ref{fig3}  for the case $m=0$, $k=0$.
Let is call this range of $g$-values the ``weak coupling-regime''.

\begin{figure}
\centering
\epsfysize=20cm
\mbox{\epsffile{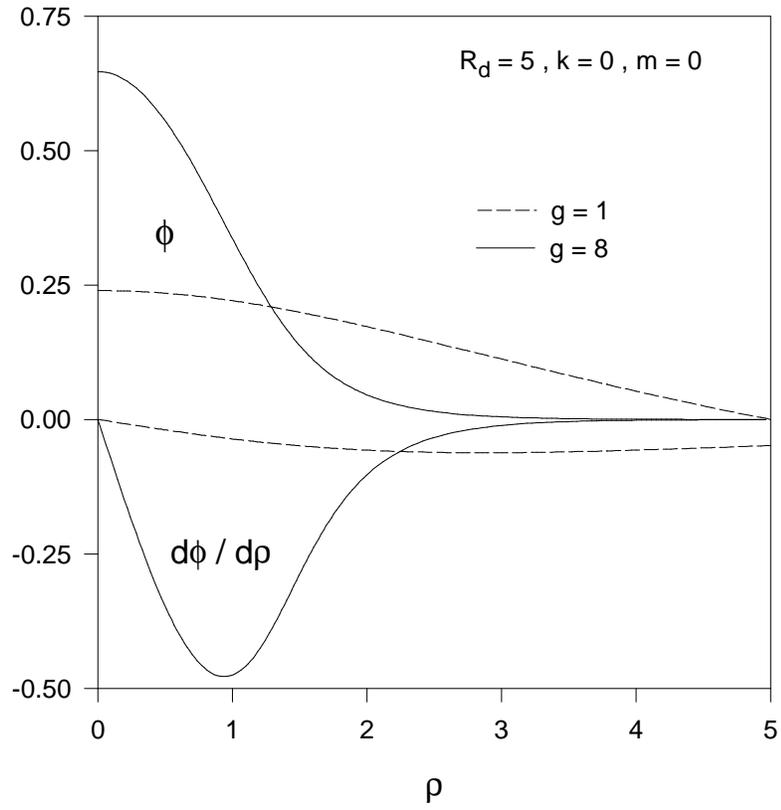}}
\caption{\label{fig4}
The profiles of the function $\phi(\rho)$ and the derivative $\frac{d\phi}
{d\rho}$
are shown for non-spinning ($m=0$), nodeless ($k=0$) solutions on the disc with radius 
$R_d=5$ for two different
values of $g$.}
\end{figure}

When $g$ is large enough,
the value of $\omega$ becomes negative (see Fig.\ref{fig3})
and the solutions have an 
 inflexion point at some intermediate value of $\rho$
(see Fig.\ref{fig4}). As a consequence  they
become much more localized in the region around the origin. In this
``strong coupling-regime'' the solutions can thus be described as being
soliton-like. We also note that 
the value of  $\omega$ decreases very strongly
 with $g$.

The difference between the strong and weak coupling is
shown in Fig.\ref{fig4}
for $g=1$, respectively $g=8$ and $R_d=5$.
The value $\omega$ is shown in dependence on $g$ for the non-spinning
($m=0$) and nodeless ($k=0$) solution for different 
values of the disc-radius
$R_d$ in Fig.\ref{fig3}.

In Fig. \ref{fig3} we also superposed the values of
$\omega$ corresponding to the case of the full plane 
(i.e. $R_d=\infty$).
As has been shown in \cite{hz,bhz}, in this case, solutions exists only for 
large enough
values of $g$: $g \geq \ g_{cr} \approx 2.94$.
It is therefore natural  to try to understand the
pattern available on the plane as the $R_d \rightarrow \infty$
limit of the pattern of solutions for the disc.
In this respect,
it is interesting to notice  that in the ``strong coupling-regime''
the values of $\omega$ corresponding to a finite disc get very close
(especially for the case corresponding
to $R_d = 10$)  to the
 values of $\omega$ associated with the case of the infinite disc.
This is related to the fact that with increasing $g$, the ``soliton'' becomes
more and more peaked around the origin and doesn't ``notice'' whether the disc is
finite or infinite. 
This observation gives a natural explanation for the fact that
solutions exist only for
$g \geq \ g_{cr}$ in the case of the plane.
Indeed in the case of the plane, it was noticed \cite{bhz}
that  the linearized equation (valid asymptotically)
also is of Bessel type.
However, the nodeless, localized solutions
are exponentially decreasing and can only be associated
with the modified
Bessel function, which correspond to negative values of $\omega$. That's why,
on the plane, nodeless soliton-like solutions exist only
 for $g \geq 2.94$ and  have no ``weak coupling''-counterparts. For large
 values of $g$ the numerical results strongly suggest the following
 linear relation between $g$ and the frequency corresponding to the
 fundamental mode~:
 $\omega \approx -0.66 g + 2.19$.
 We also studied  the solutions with $k >0$ nodes
 and $m>0$ and found that the pattern is similar to the
  case $k=0$ and $m=0$ discussed above in detail.

 We do not find it useful to discuss them here, but we believe
   that our results provide a robust evidence that regular solutions
   of (\ref{eqdisc})
   with an arbitrary number of nodes and arbitrary (integer) angular
   momentum exist on the disc/plane.

%%%%%%%%%%%%%%%%%%%%%%%%%%%%%%%%%%%%%%%%%%%%%%%%%%%%%%%%%
\section{Equation on the cylinder}%%%%%%%%%%%%%%%%%%%%%%
%%%%%%%%%%%%%%%%%%%%%%%%%%%%%%%%%%%%%%%%%%%%%%%%%%%%%%%%%
We also considered the equation on a cylinder of radius
$R_c$ and length $2L$.
This is a natural extension of the previous results in order to describe
soliton-like structures on nanotubes with radius $R_c$ and length $2L$.
Using the natural coordinates
$z$, $\varphi$,  the Laplacian reads:
\begin{equation}
\Delta_{cylinder}=\frac{1}{R_c^2}
\frac{\partial^2}{\partial \varphi^2}+
\frac{\partial^2}{\partial z^2}
\end{equation}
with $\varphi\in [0:2\pi]$ and $z \in [-L,L]$. The norm reads:
\begin{equation}
\eta^2=\int\limits_0^{2\pi} 
\int\limits_{-L}^{L} |\psi|^2  dz d\varphi \
\end{equation}
and the energy is given by:
 \begin{equation}
 E=\int\limits_0^{2\pi}\int
 \limits_{-L}^{L} \left( |\vec{\nabla}\psi|^2-2g |\psi|^4 +
\frac{1}{4}g(\vec\nabla |\psi|^2)^2\right) dz d\varphi \ .
 \end{equation}

%%%%%%%%%%%%%%%%%%%%%%%%%%%%%%%%%%%%%%%%%
\subsection{Discussion of the solutions}
%%%%%%%%%%%%%%%%%%%%%%%%%%%%%%%%%%%%%%%%% 
We use the following Ansatz:
\begin{equation}
\psi(t,z,\varphi) = e^{-i \omega t + i m \varphi} F(z)   \ \ , \ \
m=0,1,2, \dots
\end{equation}
The equation (\ref{cnls}) then becomes:
\begin{equation}
\label{equacyl}
  F'' + (F')^2 \frac{gF}{1 + g F^2}
  + \left(4 g F^2 + \omega - \frac{m^2}{R_c^2}\right) \frac{F}{1 + g F^2}=0 \ ,
\end{equation}
where the prime denotes the derivative with respect to $z$.

Again, we can rewrite this equation in terms of a ``spectral equation'':
\begin{equation}
-\left[F''+(F')^2 \frac{g F}{1+gF^2}
+ \left( g F^2(4-\omega) - \frac{m^2}{R_c^2}\right) \frac{F}{1+g F^2}\right] =\omega F \ .
\end{equation}
This equation has to be solved subject to the following boundary conditions:
\begin{equation}
\label{bccy}
F(-L) = F(L) = 0 \ .
\end{equation}
The equation (\ref{equacyl}) is then solved subject to the boundary conditions
(\ref{bccy}) using the numerical routine described in \cite{colsys}.

%%%%%%%%%%%%%%%%%%%%%%%%%%%%%%%%
\subsubsection{The case $g=0$}
%%%%%%%%%%%%%%%%%%%%%%%%%%%%%%%%
The case $g=0$ corresponds to an harmonic equation:
\begin{equation}
F''=-\left(\omega-\frac{m^2}{R_c^2}\right)F  \ .
\end{equation}
The  solutions of this equation subject to the boundary conditions (\ref{bccy})
are:
\begin{equation}
F(z) = \sin\left((z-L) \sqrt{\omega - m^2/R_c^2}\right) \ \ \ {\rm with} \ \ \
\omega = \left(\frac{m}{R_c}\right)^2 +
\left(\frac{\pi n}{2 L} \right)^2
\ \ , \ \ n = 1,2,3,\dots
\end{equation}
\subsection{The case $g \neq 0$}
It is possible to solve (\ref{equacyl}) by quadrature, which gives:
\begin{equation}
(F')^2 = \frac{C-2 g F^4 - \Omega F^2}{1+gF^2} \ \ , \ \
\Omega \equiv \omega - \frac{m^2}{R_c^2}
\end{equation}
and $C$ is an integration constant. The integration can
be performed, however, it leads to an involved and untractable expression
of $z$ in terms of $F$, $C$, $\Omega$. Assuming the cylinder to be
infinite ($L= \infty$) and $F(\infty)= F'(\infty) = 0$ leads to $C=0$.
If we now require in addition that the solution is symmetric under the
reflexion $z \rightarrow -z$ (i.e. $F'(0) = 0$) we obtain the 
relation $\Omega = - 2 g F(0)^2$. This provides a useful
check for  our numerical results.

Performing a numerical analysis of the values of $\omega$ as a 
function of  $g$ (with $L$, $R_c$, $k$, $m$ fixed), we obtain a pattern
very similar to that available for the case on a disc
(see Fig.\ref{fig3}). The only noticable difference resides in the
fact that, in the limit $L = \infty$, a normalizable solution
can be constructed for all values of the parameter $g$. 
This case is, of course, interesting with iew to nanotubes since
the lengths $2L$ of a tube is much larger than its diameter $2R_c$.
In the case
of the ``ground state'' solution, which is non-spinning ($m=0$) and
nodeless ($k=0$),
the following relations hold :
\begin{equation}
 \omega \approx - 0.1 g^2  \ \ {\rm for} \  \ g \ll 1  \ \ \ , \ \ \
\omega \approx -0.555 g + 1.28 \ \ {\rm for} \  \ g \gg 1  \ .
\end{equation}
Unlike in the case of the sphere,
we did not succeed in constructing asymmetric solutions in the case
of the cylinder (i.e. solutions, which are neither even nor odd
under the reflexion $z \rightarrow -z$) and we believe that there
are no such solutions.

\begin{equation}
\end{equation}

\section{Conclusions}
In this paper we have presented an extended analysis
of a modified non-linear Schr\"odinger equation in 1+2 dimensions.
This equation was constructed as an effective continuous equation
describing an interaction of an excitation with a 2-dimensional
hexagonal lattice. In this paper, we think of the excitation
as an ``excess electron'' described by a wave function $\psi$ and the
interaction   to be ``electron-phonon'' -like. As is well-known from amide I-excitations
in biopolymeric lattices such an interaction leads to the creation of a localised state,
a ``soliton''.

The continuous modified non-linear Schr\"odinger
equation was  obtained with the idea 
to approximate the physical system  on the 
plane \cite{hz,bhz}, but it can be modified in order to
be considered on different geometrical manifolds, which are
justified physically. These are, namely, a cylinder 
to mimick excitations on a carbon nanotube
and a sphere, respectively, in order to describe excitations
(e.g. extra electrons) on a fullerene
nanomolecule, which is particularly interesting
with view to the possibility of describing the high transition
temperature of superconducting fullerides.

Completed with the different types of
boundary conditions related to the geometry and using an Ansatz previously introduced
in the study of boson stars \cite{bs} and spinnning Q-balls \cite{volkov},
the equation (\ref{cnls}) possesses a lot symmetries, some are
continuous  (like rotations), some are discrete
(like reflexions).

In all cases considered the spectrum of solutions is rich, many
``normal-mode''-types of solutions exist, mainly characterized by two
integers: the internal angular momentum (or spin) and the
number of nodes of the wave function. This is very reminiscent to the
principal quantum numbers of more familiar quantum mechanical systems.

It might turn out to be difficult to  classify the solutions of a 
non-linear, partial differential equations. The
various types of solutions, however, can be
traced back from 
the corresponding solutions
available in the weak coupling limit,
in which the equation becomes linear and its spectrum is known analytically.
In the case of the sphere, disc and cylinder, respectively,
we found deformed Legendre functions, deformed Bessel functions
and deformed harmonic functions.
However, in the case of the sphere, we demonstrated the existence of
additional solutions, which are specifically related to the non-linear
character of the equation.
The main feature distinguishing these supplementary
solutions is their asymmetry with respect to the reflexion at
the equator of the sphere. In other words, we exhibited
the spontaneous symmetry breaking (SSB)
of one of the natural symmetries of the initial problem.

This SSB  occurs when the non-linear coupling becomes strong enough.
Similarly it can occur for very different types of non-linear
equations. As an example, we point out the bifurcation
of bisphaleron solutions from the sphaleron solutions available
in the case of the classical
equations of the electroweak field theory \cite{kb,yaffe}. In this 
case, the parity operator is spontaneously broken by the so called
bisphaleron solutions.\\
\\
\\
{\bf Acknowledgments} Y.B. gratefully acknowledges the Belgian F.N.R.S.
for financial support.

\newpage

\newpage

\newpage

\newpage
 \end{document}